\documentstyle[12pt]{article}
\begin{document}

\title{ Renormalization-Group Effects \\ in the SO(3) Gauge Model With Nearly Degenerate Neutrino 
Masses and Bi-maximal Mixing }
\author{Yue-Liang  Wu\footnote{YLWU@ITP.AC.CN}  \\  
Institute of Theoretical Physics, Chinese Academy of Sciences \\
Beijing 100080, P.R. China } 
\date{AS-ITP-99-11, hep-ph/9905222}
\maketitle
\begin{abstract} 
The renormalization-group effects on the scenario with nearly degenerate mass and bi-maximal mixing 
of neutrinos are analyzed. As a consequence, we explicitly show in the SO(3) gauge model that 
the renormalization-group effects could modify the fitting values of the relevant basic parameters 
of the model, but leave the nearly bi-maximal mixing scenario to remain stable for reconciling both 
solar and atmospheric neutrino data. The approximately degenerate Majorana neutrino masses are allowed 
to be large enough to play a significant cosmological role without conflicting with the current 
neutrinoless double beta decay.
\end{abstract}
{\bf PACS numbers: 12.15F, 11.30H} \\

\newpage

   Evidences for oscillation of atmospheric neutrinos \cite{SUPERK1} and for deficit of 
solar neutrinos\cite{SUPERK2} provide strong indications that neutrinos are massive. 
If there exist only three light neutrinos, the mixing matrix is likely to be nearly bi-maximal 
with approximately degenerate masses \cite{DNM,GGV,YLW,SO} in order to simultaneously satisfy the 
requirements that both solar and atmospheric neutrino flux anomalous are explained by neutrino 
oscillations, and the evolution of the large-scale structure of the universe\cite{HDM} gets significant
contributions from massive neutrinos. It has also been noticed \cite{GGV} that these remarkable 
features will result in a nice form for neutrino mass matrix in the weak gauge and charged lepton 
mass basis. Such a nice form of the neutrino mass matrix could be generated by using continuous 
or discrete symmetries\cite{YLW,SO}. In ref. \cite{YLW}, we have shown that a realistic scheme with nearly 
degenerate neutrino masses and bi-maximal mixing can be derived in the simple extension 
of the standard model with gauged SO(3) lepton flavor symmetry. The mass-squared 
difference $\Delta m_{sun}^{2}$ required from solar neutrinos is usually much smaller than the one 
$\Delta m_{atm}^{2}$ from the atmospheric neutrinos, i.e., $\Delta m_{sun}^{2}\ll 
\Delta m_{atm}^{2}$. Numerically, the values of $\Delta m_{sun}^{2}$ are of order 
$\Delta m_{sun}^{2} \sim 10^{-4}-10^{-5} eV^{2}$ for matter enhanced  
MSW solutions\cite{MSW} with large mixing angle\cite{LAMSW}, and of order $10^{-10}-10^{-11} eV^{2}$ 
for vacuum oscillation solutions\cite{VO1,VO2}. In such a situation, the renormalization group effects
must be considered. Recently, it has been shown\cite{EL} that, for the fixed pattern of neutrino mixing, 
the bi-maximal mixing is not stable due to renormalization group effects, 
and the renormalized masses and mixing angles are not compatible 
with all the experimental constraints. The same subject has also been studied in \cite{CEIN}, 
similar (negative) conclusions have been reached, however, it has been 
found that when the general see-saw scenario as the mechanism responsible for
generating the effective neutrino mass matrix, some positive results could be obtained in large regions 
of the parameter space consistent with the large angle MSW solution.   
In this note, we will explicitly show that in the SO(3) gauge model the renormalization group effects 
could modify the fitting values of the relevant basic parameters of the model, but the scenario 
with nearly degenerate mass and bi-maximal mixing of neutrinos could remain stable to  
accommodate both solar and atmospheric neutrino data, and the neutrino masses could still
be large enough to play a significant cosmological role without conflicting with the 
current neutrinoless double beta decays. 
  
  We start from a general consideration. Let $M_{e}$ and $M_{\nu}$ are mass matrices of charged-leptons
and neutrinos respectively, they are defined as ${\cal L}_{m} = \bar{e}_{L}M_{e}e_{R} + 
\bar{\nu}_{L} M_{\nu} \nu_{L}^{c} + H.c. $. $M_{e}$ can always be written as 
$M_{e} = U_{e} D_{e} V_{e}^{\dagger}$ with $U_{e}$ and $V_{e}$ the two unitary matrices and 
$D_{e} = diag. (m_{e}. m_{\mu}, m_{\tau})$ the diagonal mass matrix for the charged leptons. When 
the CKM-type lepton mixing matrix is mainly determined from the charged lepton mass matrix, i.e., 
the unitary matrix $U_{e}^{\dagger}$, and the neutrino masses are nearly degenerate, 
then the neutrino mass matrix can be written as  
\begin{equation}
M_{\nu} = m_{0} \left( I_{3} + \epsilon_{\nu} S_{\nu} \right) 
\end{equation}
where $I_{3}$ is an $3\times 3$ unit matrix, $m_{0}$ is a mass parameter with $m_{0} = O(1)$ eV
 and $S_{\nu}$ is a symmetric matrix with matrix elements 
being order of one. Thus $\epsilon_{\nu}$ is a small paramter $\epsilon_{\nu}\ll 1$. 
   
   In the weak gauge and charged-lepton mass basis, the neutrino mass matrix is given by   
\begin{equation}
\tilde{M}_{\nu} \equiv U_{e}^{\dagger} M_{\nu} U_{e} = m_{0}\left( U_{e}^{\dagger}U_{e}^{\ast} + 
\epsilon_{\nu} U_{e}^{\dagger}S_{\nu}U_{e}^{\ast} \right) 
\end{equation}
In this basis, when the renormalization group equations evaluate from the high energy scale $\Lambda$ to the 
electroweak scale, the renormalized neutrino mass matrix is simply given by
\begin{eqnarray}
M_{\nu}^{R} & = &  R\tilde{M}_{\nu}R = \hat{m}_{0} \{ U_{e}^{\dagger}U_{e}^{\ast} + 
\epsilon_{\nu} U_{e}^{\dagger}S_{\nu}U_{e}^{\ast} 
- \epsilon_{R} D U_{e}^{\dagger}U_{e}^{\ast} 
- \epsilon_{R} U_{e}^{\dagger}U_{e}^{\ast} D  \nonumber \\
& - &  \epsilon_{R}\epsilon_{\nu} D U_{e}^{\dagger}S_{\nu}U_{e}^{\ast}  
- \epsilon_{R}\epsilon_{\nu}   U_{e}^{\dagger}S_{\nu}U_{e}^{\ast} D  \\
& + &  \epsilon_{R}^{2} D U_{e}^{\dagger}U_{e}^{\ast} D 
+ \epsilon_{R}^{2}\epsilon_{\nu} D U_{e}^{\dagger}S_{\nu}U_{e}^{\ast} D  \} \nonumber 
\end{eqnarray}   
with $R = diag. (R_{e}, R_{\mu}, R_{\tau}) 
\equiv R_{e}(I_{3} + \epsilon_{R} D)$ and 
\begin{equation}
D = \left( \begin{array}{ccc}
   0 & 0 & 0  \\
   0 & \epsilon & 0 \\
 0 & 0 &  1 \\ 
\end{array} \right) 
\end{equation}
where $\epsilon_{R} = 1- R_{\tau}/R_{e}$,  
$\epsilon = (1-R_{\mu}/R_{e})/\epsilon_{R}$ and $\hat{m}_{0} = R_{e}m_{0}$. 
$R_{f}$ ($f=e, \mu, \tau$) are the Yukawa renormalization factors\cite{EL}  
and defined as 
\begin{equation}
R_{f} = exp [ \frac{1}{16\pi^{2}} \int_{t_{0}}^{t} h_{f}^{2} dt ]
\end{equation}
with $t = \ln \mu$ and $h_{f}$ ($f=e, \mu, \tau$) being the Yukawa coupling constants of 
the charged leptons.  By a unitary transformation $\nu_{L} \rightarrow U_{e}^{\dagger} \nu_{L}$, 
the mass matrix becomes 
\begin{eqnarray}
\hat{M}_{\nu} & \equiv & U_{e}M_{\nu}^{R}U_{e}^{T} = \hat{m}_{0}\{ I_{3} + 
\epsilon_{\nu} S_{\nu} 
- \epsilon_{R} (\hat{D} + \hat{D}^{\ast} ) \nonumber \\
& - &   \epsilon_{R}\epsilon_{\nu} ( \hat{D} S_{\nu} + S_{\nu}\hat{D}^{\ast})  
+  \epsilon_{R}^{2} \hat{D}\hat{D}^{\ast} 
+ \epsilon_{R}^{2}\epsilon_{\nu} \hat{D} S_{\nu} \hat{D}^{\ast}  \} 
\end{eqnarray}
with $\hat{D} = U_{e} D U_{e}^{\dagger}$. In general, the resulting value for $\epsilon_{R}$ and 
$\epsilon$ are small, i.e., $\epsilon_{R}\ll 1$ and $\epsilon\ll 1$. 
Thus the neutrino masses should remain almost degenerate, but the 
fitting values of the relevant parameters appearing in the matrix $S_{\nu}$ have to be modified. 
To be manifest, we shall consider an explicit scenario discussed in our recent papers \cite{YLW}. 
From there we can read 
\begin{equation}
U_{e}^{\dagger}\simeq \left( \begin{array}{ccc}
  ic_{1} & -s_{1} & 0  \\
 c_{2}s_{1} & -ic_{1}c_{2} & -s_{2} \\
 s_{1} s_{2} & -ic_{1}s_{2} & c_{2}  \\
\end{array} \right)
\end{equation}
and 
\begin{equation}
S_{\nu} = \left( \begin{array}{ccc}
  r_{\Delta} s_{1}^{2} & 0 & 2r_{\delta}s_{2}c_{2}  \\
  0 & r_{\Delta} c_{1}^{2} &  0 \\
  2r_{\delta} s_{2}c_{2} & 0
 &  1  \\ 
\end{array} \right)
\end{equation}
with $\epsilon_{\nu} \equiv \Delta_{+}$, $r_{\Delta} \equiv \Delta_{-}/\Delta_{+}$ and 
$r_{\delta} \equiv \delta_{-}/\Delta_{+}$. Here $\Delta_{\pm}= \delta_{+}\pm \delta_{-}\cos2\theta_{2}$ 
with $\delta_{+}$ and $\delta_{-}$  the two basic parameters of the model. 
It is not difficult to check that the matrix $\hat{D} = U_{e} D U_{e}^{\dagger}$ has the following
explicit form
\begin{equation}
\hat{D} = \left( \begin{array}{ccc}
  \hat{s}_{2}^{2}s_{1}^{2} & i\hat{s}_{2}^{2}c_{1}s_{1} & rs_{1}c_{2}s_{2}  \\
 i\hat{s}_{2}^{2}c_{1}s_{1}  & -\hat{s}_{2}^{2}c_{1}^{2} & irc_{1}c_{2}s_{2} \\
 rs_{1}c_{2}s_{2} & irc_{1}c_{2}s_{2} & \hat{s}_{2}^{2}  \\
\end{array} \right)
\end{equation}
with $\hat{s}_{2}^{2} = s_{2}^{2} + \epsilon c_{2}^{2}$ and $r = 1- \epsilon $. 
It is seen that to the linear terms of $\epsilon_{\nu}$ and $\epsilon_{R}$ the neutrino mass matrix  
$\hat{M}_{\nu}$ has the same structure as $M_{\nu}$, thus the mass matrix $\hat{M}_{\nu}$ 
can be written in the following form
\begin{equation}
\hat{M}_{\nu} = \hat{m}_{0}\{ I_{3} + 
\hat{\epsilon}_{\nu} \hat{S}_{\nu}  
-   \epsilon_{R}\epsilon_{\nu} ( \hat{D} S_{\nu} + S_{\nu}\hat{D}^{\ast})  
+  \epsilon_{R}^{2} \hat{D}\hat{D}^{\ast} 
+ \epsilon_{R}^{2}\epsilon_{\nu} \hat{D} S_{\nu} \hat{D}^{\ast}  \} 
\end{equation} 
with 
\begin{equation}
\hat{S}_{\nu} = \left( \begin{array}{ccc}
  \hat{r}_{\Delta} s_{1}^{2} & 0 & 2 \hat{r}_{\delta}s_{2}c_{2}  \\
  0 & \hat{r}_{\Delta} c_{1}^{2} &  0 \\
  2 \hat{r}_{\delta} s_{2}c_{2} & 0
 &  1  \\ 
\end{array} \right)
\end{equation}
and $\hat{\epsilon}_{\nu} \equiv \hat{\Delta}_{+}$, $\hat{r}_{\Delta} \equiv 
\hat{\Delta}_{-}/\hat{\Delta}_{+}$, $\hat{r}_{\delta} \equiv \hat{\delta}_{-}/\hat{\Delta}_{+}$.
Here $\hat{\Delta}_{\pm} = \Delta_{\pm}- \epsilon_{R} \hat{s}_{2}^{2}$ and 
$\hat{\delta}_{-}=\delta_{-} - r s_{1} \epsilon_{R}$. 

     It is seen that when only keeping to the first order of $\epsilon_{\nu}$ and $\epsilon_{R}$, 
the scenario with nearly degenerate neutrino masses and bi-maximal mixing is automatically preserved. 
The only difference is that one shall use the modified parameters  $\hat{\Delta}_{\pm} = 
\Delta_{\pm}- \epsilon_{R} \hat{s}_{2}^{2}$ and $\hat{\delta}_{-}=\delta_{-} - r s_{1} \epsilon_{R}$, 
instead of the parameters  $\Delta_{\pm}$ and $\delta_{-}$, to accommodate the experimental data. 
The resulting constraints in this approximation are given by
\begin{equation}
\hat{\Delta}_{+} - \hat{\Delta}_{-}c_{1}^{2} \simeq \Delta m_{atm}^{2} /2m_{0}^{2} 
\end{equation}
and 
\begin{equation}
|\hat{\delta}_{-}\sin2\theta_{2}| \ll |\hat{\Delta}_{+} - \hat{\Delta}_{-}s_{1}^{2}| 
\end{equation}
Considering $s_{2}\simeq c_{2}$ and $s_{1}\simeq c_{1}$, 
these constraints can be simplified to be 
\begin{equation}
\delta_{+} - \epsilon_{R}\hat{s}_{2}^{2} \simeq \Delta m_{atm}^{2} /m_{0}^{2} 
\end{equation}
and 
\begin{equation}
c_{1}^{2}-s_{1}^{2} + \left(\frac{\delta_{-} - r\epsilon_{R}s_{1}}{(\delta_{+} 
-\epsilon_{R}\hat{s}_{2}^{2})c_{1}}\right)^{2} \simeq \frac{\Delta m_{sun}^{2}}{2\Delta m_{atm}^{2}} 
\end{equation}
with $m_{0} \simeq O(1)$ eV.  The question is how the higher order terms affect the stability of 
the scenario. For convenience of discussions, we may change the basis by a phase redifinition 
$e_{L} \rightarrow i e_{L}$ and $\nu_{\mu L} \rightarrow i \nu_{\mu L} $, so that the lepton 
mixing matrix becomes orthogonal 
\begin{equation}
U_{e}^{\dagger}\rightarrow O_{e}^{T} \simeq \left( \begin{array}{ccc}
  c_{1} & s_{1} & 0  \\
 c_{2}s_{1} & -c_{1}c_{2} & -s_{2} \\
 s_{1} s_{2} & -c_{1}s_{2} & c_{2}  \\
\end{array} \right)
\end{equation}
and the neutrino mass matrix gets the following general form 
\begin{equation}
\hat{M}_{\nu} = m_{0}\left( \begin{array}{ccc} 
  1 + \hat{\epsilon}_{\nu}\hat{r}_{\Delta} s_{1}^{2} + a_{1}\epsilon_{R}^{2} & b_{1}\epsilon_{R}^{2} 
& 2 \hat{\epsilon}_{\nu} \hat{r}_{\delta}s_{2}c_{2} + b_{2}\epsilon_{R}^{2}  \\
  b_{1}\epsilon_{R}^{2} & -1 - \hat{\epsilon}_{\nu}\hat{r}_{\Delta} c_{1}^{2} - a_{2}\epsilon_{R}^{2} & 
 b_{3}\epsilon_{R}^{2} \\
  2 \hat{\epsilon}_{\nu}\hat{r}_{\delta} s_{2}c_{2} + b_{2}\epsilon_{R}^{2} & b_{3}\epsilon_{R}^{2}
 &  1 + \hat{\epsilon}_{\nu} + a_{3}\epsilon_{R}^{2} \\ 
\end{array} \right)
\end{equation}
where the coefficients $a_{i}$ and $b_{i}$ (i=1,2,3) are given by $s_{1}$ and $s_{2}$, and contain 
higher order terms of $\epsilon_{R}$. Before going to a detailed discussion, we consider the following 
$2\times 2$ matrix 
\begin{equation}
M \simeq m_{0}\left( \begin{array}{cc}
  -1 - \epsilon_{1} & \epsilon_{2}  \\
 \epsilon_{2} & 1 + \epsilon_{3}
\end{array} \right)
\end{equation}
which is diagonalized by an $2\times 2$ rotational matrix. The rotational angle $\theta$ is given by 
\begin{equation}
\tan 2 \theta = 2\epsilon_{2}/(2 + \epsilon_{1} + \epsilon_{3})
\end{equation}
The eigenvalues are 
\begin{eqnarray}
m_{1} & = &  m_{0} \left(-\sqrt{1+ \epsilon_{1} + \epsilon_{3} + \epsilon_{2}^{2} + 
(\epsilon_{3}-\epsilon_{1})^{2}/4 } + (\epsilon_{3}-\epsilon_{1})/2 \right) \nonumber \\
m_{2} & = &  m_{0} \left(\sqrt{1+ \epsilon_{1} + \epsilon_{3} + \epsilon_{2}^{2} + 
(\epsilon_{3}-\epsilon_{1})^{2}/4 } + (\epsilon_{3}-\epsilon_{1})/2 \right)
\end{eqnarray}
 The mass-squared difference is propotional to $(\epsilon_{3} - \epsilon_{1})$
\begin{equation}
\Delta m^{2} \equiv m_{2}^{2} -m_{1}^{2} = 2m_{0}^{2}(\epsilon_{3} - \epsilon_{1})
\sqrt{1+ \epsilon_{1} + \epsilon_{3} + \epsilon_{2}^{2} + 
(\epsilon_{3}-\epsilon_{1})^{2}/4 } 
\end{equation}
when $|\epsilon_{i}|\ll 1$, one has 
\begin{equation}
\Delta m^{2} \equiv m_{2}^{2} -m_{1}^{2} \simeq 2m_{0}(M_{22}+M_{11})
\end{equation}
Obviously, when $\epsilon_{1} = \epsilon_{3}$, , i.e., $M_{11} = - M_{22}$, 
one has $m_{1} = - m_{2}$ and $\Delta m^{2} \equiv m_{2}^{2} -m_{1}^{2} = 0$. 
Noticing this feature, we now return to discuss the realistic neutrino 
mass matrix $\hat{M}_{\nu}$. As the first step, we transfer the `23' texture into zero via 
an orthogonal transformation $O_{23}$,  
\begin{equation}
O_{23} = \left( \begin{array}{ccc}
  1 & 0 & 0  \\
 0 & c'_{2} & s'_{2} \\
 0 & -s'_{2} & c'_{2}  \\
\end{array} \right)
\end{equation}
with 
\begin{equation}
\tan2\theta'_{2} = \frac{2b_{3}\epsilon_{R}^{2}}{2 +\epsilon_{\nu}(1+\hat{r}_{\Delta} c_{1}^{2})
 + (a_{2} + a_{3})\epsilon_{R}^{2} }
\end{equation}
so that the mass matrix $\hat{M}'_{\nu}\equiv O^{T}_{23}\hat{M}_{\nu}O_{23} $ has the following form
\begin{equation}
\hat{M}'_{\nu} = m_{0}\left( \begin{array}{ccc}
  1 + \hat{\epsilon}_{\nu}\hat{r}_{\Delta} s_{1}^{2} + a'_{1}\epsilon_{R}^{2} & b'_{1}\epsilon_{R}^{2} 
& 2 \hat{\epsilon}_{\nu} \hat{r}_{\delta}s_{2}c_{2} + b'_{2}\epsilon_{R}^{2}  \\
  b'_{1}\epsilon_{R}^{2} & -1 - \hat{\epsilon}_{\nu}\hat{r}_{\Delta} c_{1}^{2} - a'_{2}\epsilon_{R}^{2} & 0 \\
  2 \hat{\epsilon}_{\nu}\hat{r}_{\delta} s_{2}c_{2} + b'_{2}\epsilon_{R}^{2} & 0
 &  1 + \hat{\epsilon}_{\nu} + a'_{3}\epsilon_{R}^{2} \\ 
\end{array} \right)
\end{equation}
with $a'_{i} = a_{i} + \alpha_{i}\epsilon_{R}^{2}$ and $b'_{i} = b_{i} 
+ \beta_{i}\epsilon_{R}^{2}$ (i=1,3). Where $\alpha_{i}$ and $\beta_{i}$ are relevant to 
the angle $\theta'_{2}$. To accommodate the experimental data, a simple treatment is to set 
\begin{eqnarray}
& &  (\hat{M}'_{\nu})_{13} = m_{0}\left( 2 \epsilon_{\nu} \hat{r}_{\delta}s_{2}c_{2} 
+ b'_{2}\epsilon_{R}^{2}\right) = 0   \nonumber \\
& & (\hat{M})_{33}^{2} - (\hat{M})_{22}^{2} = 2m_{0}^{2} [ (1- \hat{r}_{\Delta} c_{1}^{2})
\hat{\epsilon}_{\nu} 
+ (a'_{3}-a'_{2}) \epsilon_{R}^{2} ] \simeq \Delta m_{atm}^{2} \\
& & (\hat{M}'_{\nu})_{11} + (\hat{M}'_{\nu})_{22} = m_{0}   
 [(s_{1}^{2}-c_{1}^{2})\hat{\epsilon}_{\nu}\hat{r}_{\Delta} + (a'_{1}-a'_{2})\epsilon_{R}^{2} ]
\simeq -\Delta m_{sun}^{2}/2m_{0}  \nonumber 
\end{eqnarray}
By an explicit evaluation, we arrive at the following constraints
\begin{equation}
\delta_{-} = rs_{1}\epsilon_{R} + O(\epsilon_{R}^{2}) 
\end{equation}
and
\begin{eqnarray}
& & \delta_{+}[1-2\hat{s}_{2}^{2}(1+c_{1}^{2})\epsilon_{R}]s_{1}^{2}-
 \hat{s}_{2}^{2}s_{1}^{2}\epsilon_{R}  = \Delta m_{atm}^{2} /2m_{0}^{2} \nonumber \\
& & \qquad -\delta_{-} \{ [1 + c_{1}^{2} + 2(1 + c_{1}^{4})\hat{s}_{2}^{2}\epsilon_{R} ](c_{2}^{2}-s_{2}^{2}) 
-rs_{1}\sin 2\theta_{2} \epsilon_{R} \}  \nonumber \\
& & \qquad -[\hat{s}_{2}^{2} + 0.25r^{2}c_{1}^{2}\sin2\theta_{2} + (0.25r^{2}\sin 2\theta_{2} 
+ \hat{s}_{2}^{2}c_{1}^{2})(s_{1}^{2} - c_{1}^{2}) ] \epsilon_{R}^{2} + O(\epsilon_{R}^{4})  \\
& & [\delta_{+} (1 - 2\hat{s}_{2}^{2}\epsilon_{R} ) - \hat{s}_{2}^{2} \epsilon_{R}  
- \delta_{-}(1 - 2\hat{s}_{2}^{2}\epsilon_{R} )(c_{2}^{2} - s_{2}^{2}) 
+ \hat{s}_{2}^{2} \epsilon_{R}^{2}](s_{1}^{2} - c_{1}^{2}) \nonumber \\
& & \qquad  = (\delta_{-}s_{1} - 0.25 r\epsilon_{R}) r \sin 2\theta_{2} \epsilon_{R} 
- \Delta m_{sun}^{2} /2m_{0}^{2} + O(\epsilon_{R}^{4}) 
\end{eqnarray}
Considering the maximal mixing suggested from the atmospheric neutrino oscillation, i.e., 
$\theta_{2} = \pi/4$, the above constraints are simplified to be
\begin{equation}
\delta_{+} = \frac{\Delta m_{atm}^{2}}{2s_{1}^{2}m_{0}^{2}}  + \epsilon_{R}/2 
+ \frac{c_{1}^{2}\epsilon_{R}^{2}/2}{1-(1+c_{1}^{2}) \epsilon_{R}} 
+ \frac{c_{1}^{2}}{4s_{1}^{2}}\epsilon_{R}^{2} + O(\epsilon_{R}^{3})
\end{equation}
and 
\begin{equation}
\{ \frac{4\Delta m_{atm}^{2}}{m_{0}^{2}} + 
2(1 + s_{1}^{2})(c_{1}^{2} - s_{1}^{2})\epsilon_{R}^{2} +  O(\epsilon_{R}^{3}) \} 
(s_{1}^{2}-c_{1}^{2}) = 2 s_{1}^{2} [\epsilon_{R}^{2} - \frac{2\Delta m_{sun}^{2} }{m_{0}^{2}} 
+ O(\epsilon_{R}^{4})] 
\end{equation}
The most recent upper limit from the neutrinoless double beta decay\cite{DB} leads the constraint on the 
mixing angle $\theta_{1}$ to be
\begin{equation}
|s_{1}^{2} - c_{1}^{2}| < 0.2eV/m_{0}
\end{equation}
which implies that for $m_{0} = O(1)$ eV the renormalization group factor $\epsilon_{R}$ 
must satisfy the following condition 
\begin{equation}
 \epsilon_{R}^{2} <  \frac{0.2 eV/m_{0}}{1 + 0.2 eV/m_{0}}\  
 \frac{4\Delta m_{atm}^{2}}{m_{0}^{2}} + \frac{2\Delta m_{sun}^{2} }{m_{0}^{2}}
\end{equation} 
For $\Delta m_{atm}^{2} \simeq 2.5 \times 10^{-3} eV^{2}$ and 
$\Delta m_{sun}^{2} \ll \Delta m_{atm}^{2}$, we have  
\begin{eqnarray}
& & \epsilon_{R}^{2} < 2.3\times 10^{-4}, \qquad for \qquad  m_{0}=2 eV, \nonumber \\
& & \epsilon_{R}^{2} < 1.7\times 10^{-3}, \qquad for \qquad  m_{0}=1 eV 
\end{eqnarray} 

    In the SO(3) gauge model, the leptons get masses after the SO(3) gauge symmetry breaking 
down at the energy scale $\Lambda \geq 10^{5}$ GeV\cite{YLW}. From the numerical values of the Yukawa 
renormalization factors calculated in ref.\cite{EL},  one sees that the renormalization group 
effects lead to $\epsilon_{R} \simeq 10^{-6}\sim 10^{-2}$ and $\epsilon \simeq 0 \sim 5\times 10^{-2}$ 
for $\tan\beta = 1\sim 60$ in the minimal supersymmetric standard model. From the table 2 in ref.\cite{EL}, 
one can read off the following results 
\begin{eqnarray}
& & \epsilon_{R}^{2} \leq 1.8\times 10^{-4}, \qquad for \qquad \tan \beta \leq 58 \nonumber \\
& & \epsilon_{R}^{2} < 10^{-9}, \qquad for \qquad \tan \beta < 10
\end{eqnarray}
which is consistent with the above constraint. 

It is seen that for large $\tan \beta$
the parameter space is reasonable for a large angle MSW solution of solar neutrinos even when 
$s_{1}^{2}=c_{1}^{2}$. For low $\tan \beta$, the above constraints can be further simplified to be 
\begin{eqnarray}
& & \delta_{-}  \simeq  \epsilon_{R}/\sqrt{2} , \nonumber \\
& & \delta_{+} \simeq \frac{\Delta m_{atm}^{2}}{m_{0}^{2}}  + \epsilon_{R}/2 \ , \\
& & s_{1}^{2} - c_{1}^{2} \simeq \frac{m_{0}^{2}\epsilon_{R}^{2}}{4\Delta m_{atm}^{2}} - 
\frac{\Delta m_{sun}^{2} }{2\Delta m_{atm}^{2}} \nonumber 
\end{eqnarray} 
When $s_{1}^{2}=c_{1}^{2}$, only vacuum oscillation solutions could be allowed at very low 
$\tan \beta$. For a large angle MSW solution, we then obtain at low $\tan \beta$ the following relation 
between the mixing angle $\theta_{1}$ and the mass-squared diffenrences
\begin{equation}
c_{1}^{2} - s_{1}^{2} \simeq \frac{\Delta m_{sun}^{2} }{2\Delta m_{atm}^{2}}
\end{equation}
which indicates that the mixing angle $\theta_{1}$ is nearly maximal.
Note that the above constraints have been obtained by 
setting $(\hat{M}'_{\nu})_{13}=0$. For $\epsilon_{R}^{2} \ll 
\Delta m_{sun}^{2}/ 2m_{0}^{2}$, the renormalization group effects become less important. 
In this situation, one can also consider the reasonable approximation presented in eqs. (12) and (13). 
Thus the allowed range of the parameter space will be given by
eqs. (14) and (15). In such a case, the mixing angle $\theta_{1}$ can be maximal. 
   
  With the above analyses,  we may come to the conclusion that when the renormalization group effects 
are considered, the realistic scenario with approximately degenerate mass and nearly bi-maximal mixing 
of neutrinos in the SO(3) gauge model can still be preserved to accommodate both solar 
and atmospheric neutrino data via vacuum oscillation solutions or large angle MSW solutions. 
As a consequence, the fitting values of the relevant basic parameters of the model will be modified.

 {\bf Ackowledgments}: This work was supported in part by the NSF of China under the 
grant No. 19625514.

%\end{references}

\end{document}